# Elastic Instabilities in Flows through Pillared Micro channels

S. De[1], J. van der Schaaf[2], N.G. Deen[3], J.A.M. Kuipers[1], E.A.J.F. Peters[1], J.T. Padding[1,*]
1 Multiphase Reactors Group, Department of Chemical Engineering and Chemistry, Eindhoven University of Technology, The Netherlands
2 Chemical Reactor Engineering Group, Department of Chemical Engineering and Chemistry, Eindhoven University of Technology, The Netherlands
3 Multiphase and Reactive Flows Group, Department of Mechanical Engineering, Eindhoven University of Technology, The Netherlands
*Corresponding author: *J.T.Padding@tue.nl*

**Abstract**

Viscoelastic fluids exhibit elastic instabilities in simple shear flow and flow through curved streamlines. Surprisingly, we found in a porous medium such fluids show strikingly different hydrodynamic instabilities depicted by very large sideways excursions and presence of fast and slow moving lanes which have not been reported before. Particle image velocimetry (PIV) measurements through a pillared microchannel, provide experimental evidence of such instabilities at very low Reynolds number (< 0.01). We observe a transition from a symmetric laminar to an asymmetric flow, which finally transforms to a nonlinear aperiodic flow with strong lateral movements. The instability is characterized by a rapid increase in spatial and temporal fluctuations of velocity components and pressure at a critical Deborah number (De). Our experiments reveal the presence of a fascinating interplay between pore space and fluid rheology.

**Introduction**

Non Newtonian fluid sometimes exhibit time dependent fluctuations in the flow fields which are reminiscent of turbulence, yet they occur at very small Reynolds numbers (Re), a phenomenon called elastic turbulence. This may be attributed to the inherent anisotropy of the polymeric fluids [1-7]. For example, an elastic instability in a cross channel flow was observed after a critical De number in the work of Poole et al. and Arratia et al. [1, 2]. The effect of obstacles on nonlinear flow instability was studied recently by Pan et al. [3]. The onset of elastic turbulence in a straight channel was reported by several researchers [4, 5, 6]. Two dimensional elastic turbulence was numerically investigated for a simple polymeric flow in the work of Berti et al. [7]. A nonlinear stability analysis was performed for a planar Couette flow of a viscoelastic fluid in the work of Morozov and Van Saarloos [8]. The onset of elastic instabilities for complex flow structures and curved streamlines was reported by Pakdel and McKinley [9]. The concept of elastic turbulence in relation with elastic instabilities for polymeric flow was observed by Groisman et al. [10]. In case of a Taylor – Couette flow for viscoelastic fluids an extra hoop stress is produced due to the radial velocity variations, as shown by Groisman et al. [11]. Efficient mixing and chaotic flow motion in microchannel for polymeric fluids in an undulated channel was also examined by Burghelea et al. [12]. Pakdel and McKinley [13] investigated viscoelastic flow in a lid driven cavity flow and reported flow instabilities. Flow instabilities for wormlike micellar solution in a periodic array of cylinders were studied by Moss et al. [14]. The onset of elastic instability in serpentile channels was studied numerically and experimentally by Zilz et al. [15]. Mckinley et al. [16] experimentally analyzed flow transition in abrupt contractions for a viscoelastic fluid. Flow instabilities in viscoelastic flows were reviewed and an attempt was made to explain these nonlinear effects [17, 18].

Although viscoelastic fluids in simple channel flows exhibit flow instabilities, the number of pore scale studies on viscoelastic flow through complex porous media is still limited. Morais et al. [19] numerically studied non-Newtonian flow in random porous media for power-law fluids, which are non-elastic. Grilli et al. [20] performed numerical simulations for viscoelastic fluid flow over an array of cylinders. Datta et al. [21] studied spatial velocity distributions of a Newtonian fluid in a three dimensional porous medium, but in general viscoelastic flow through a porous medium is more complex. Progress in microfluidic research enables us to study these intriguing flow features at length scales which are of significant importance in oil recovery, polymer processing, packed bed flows, blood flow though tissues, medicine, geology and other applications.

In this letter, we experimentally investigate the fascinating interplay of viscoelastic effects in a model porous medium using a pillared microchannel. Due to successive contraction and expansion through the pillars, the polymer molecules get elongated and relaxed continuously, leading to buildup and release of elastic stresses. We observe that after a critical Deborah number (De) the flow becomes asymmetric, but the instabilities remain localized. At higher De the viscoelastic effects becomes so strong that the flow starts to switch from one pillar lane to another. This extreme sideways motion is associated with large nonlinear, non-periodic instabilities. We also observe an increase in apparent viscosity along with the elastic turbulence which leads us to believe that these effects must be attributed to a significant extension of polymer chains. Newtonian solutions of equal (zero-shear) viscosity do not show such flow features.

**Experimental Methodology**

Micro-PIV experiments are performed in long (6.6 cm) straight microchannels, with a width and height of 1 mm and 50 μm, respectively. The model porous medium is designed by placing an array of cylindrical pillars in a stretched hexagonal pattern from the beginning to the end of the channel as shown in figure 1. The channel and cylinders are etched in silicon and coated with carbon nanofibers. The distance along the flow direction (x) of two successive pillars ($X_P$) and along the width (y) of the channel ($Y_P$) is shown in table 1 for different channels. The number of pillars along x and y direction (n, m) is 1650 and 16 respectively for channel 1. In this letter we will mostly focus our results on experiments performed in channel 1, and use other channel results for comparison. All pillars are modified with a hydrophilic coating and microchannels are fabricated using soft lithography technique.

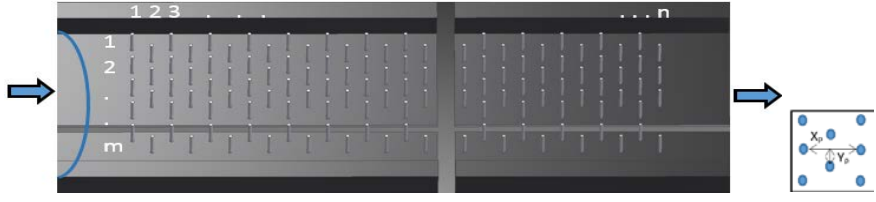

Figure 1. Geometry of a typical pillared microchannel

| Channel | Pillar diameter (μm) $(D_P)$ | X – pitch (μm) $(X_P)$ | Y- pitch (μm) $(Y_P)$ |
|---|---|---|---|
| 1 | 6 | 34 | 28.6 |
| 2 | 9 | 51 | 42.9 |
| 3 | 12 | 68 | 57.2 |

Table 1: Dimensions of different micro channels used in this study.

We investigated the flow of both Newtonian and a viscoelastic fluid through the pillared microchannel. A hydrolyzed polyacrylamide solution (HPAM, 20 MDa) is used as the non-Newtoniain fluid. The solution is prepared by adding 2000 ppm of HPAM in distilled water. The zero shear viscosity $(\eta)$ of the HPAM solution is 0.25 Pa·s and the terminal relaxation time (λ) of the polymer is 0.12 sec, as characterized by a standard strain controlled double gap rheometer at room temperature. The HPAM solution has a shear thinning rheology (see Supplementary material). Glycerol solution with similar zero shear viscosity of HPAM is used as the Newtonian fluid.

For all our experiments we have kept the Reynolds number $Re = \frac{\rho U D_P}{\eta}$ less than 0.01, so any inertial effects can be neglected. Here $U$ is the superficial flow velocity, $\rho$ is the fluid density, $\eta$ is the zero shear viscosity, and $D_P$ is the pillar diameter. The other important dimensionless number for our experiments is the Deborah number (De), which is the ratio of the relaxation time of polymer and characteristics time scale of flow. Here we will introduce two different De numbers based on length scales. The De number with regard to the pillar diameter as the characteristic length is defined as $De_P = \frac{\lambda U}{D_P}$, and De with respect to pillar-to-pillar distance (X-pitch) is defined as $De_L = \frac{\lambda U}{X_P}$. Both De numbers are relevant because the polymers experience curved and contraction-expansion flow when passing each single pillar, which has a large influence on the polymer conformation if $De_P$ is sufficiently large, while the polymers have time to relax their conformations during their flow in between the pillars if $De_L$ is sufficiently low. We note that the maximum $De_P$ reached in channel 1 is around 8.0, which is sufficiently large to trigger elastic instabilities.

In our microchannel experiments we visualize the flow characteristics for different De numbers by slowly changing the flowrate of the injected HPAM solution using a syringe pump to maintain very low flow rates. The flow visualization is done using micro Particle Image Velocimetry (μ-PIV). Fluorescent tracer particles (1 μm diameter, 0.02wt %) are seeded with the fluid. To observe the motion of the fluid inside the microchannel an inverted microscope (Zeiss optical observer) is used. The depth of field of the microscope was calculated to be 10% of the height of the microchannel. The particle tracks are visualized in a focal plane in the center between the top and bottom walls to decrease any effect of out of plane velocity gradients. The images are captured using a high speed camera mounted on the microscope. Bright field images are captured using a high intensity directed light source to excite the tracer particles. A green filter (500 – 600 nm) is used to filter any other light except the light from the particles. Micro-pressure sensors are used to measure the pressure drop accurately across the channel for different experiments. The images are collected after steady state is reached, the onset of which is ensured from the pressure signals (see Supplementary material).

Results

The temporal and spatial dependence of viscoelastic flow is studied in a square section (around 25% area of channel) close to the middle section along the channel length. Images are captured at a frame rate of 30 fps, which is much faster than the time scale of fluid flow. Figures 2 and 3 show the time averaged and spatial averaged standard deviation of velocity component along the flow direction (x) for different Deborah numbers, expressed in terms of $De_L$. In the case of time averaged analysis, for each time frame we first determine a time-dependent standard deviation $\sigma_v(t) = \sqrt{\overline{v^2(t)} - \overline{v(t)}^2}$ characterizing the difference between local velocity and an average velocity based on all velocities spatially available in the flow domain at that time (the overbar signifies spatial averaging). Then we perform a temporal averaging $\langle \sigma_v(t) \rangle$ of the obtained standard deviations for each De number (angular brackets signify temporal averaging). In case of the spatial averaging, for each point in the flow domain we first determine a spatially dependent standard deviation $\sigma_v(x,y) = \sqrt{\langle v^2(x,y) \rangle - \langle v(x,y) \rangle^2}$ characterizing the difference between temporary velocity and long-time averaged velocity at that location. Then we perform a spatial averaging $\overline{\sigma_v(x,y)}$ of the obtained standard deviations for each De number.

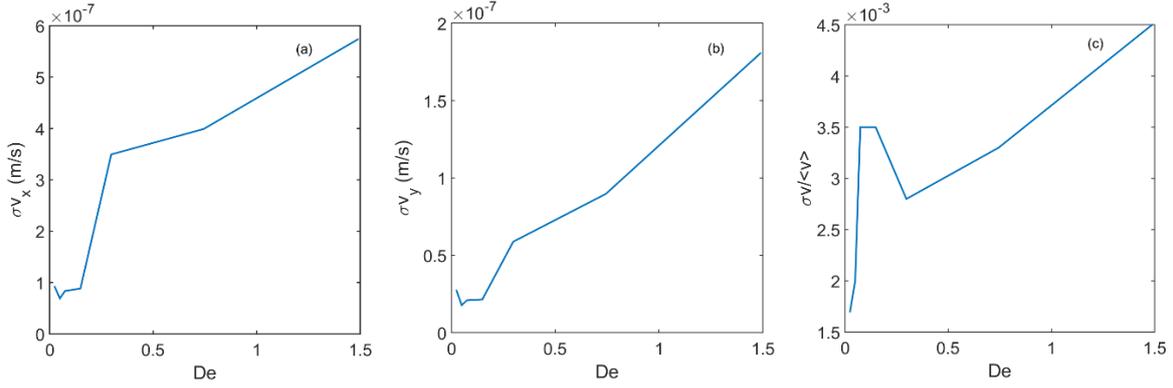

Figure 2. Time averaged velocity fluctuations vs. Deborah number $De_L$: (a) streamwise fluctuations, (b) lateral fluctuations, and (c) total fluctuating magnitude normalized by the average flow velocity

Figure 2 (a) and (b) clearly show that the time averaged velocity fluctuations increases with increasing $De_L$. We find that the velocity fluctuations along the flow direction ($\sigma_{v_x}$) are around 3 times larger than in the lateral fluctuations $\sigma_{v_y}$ and that both components sharply increase after $De_L$ around 0.25. Figure 2 (c) shows that the time averaged velocity fluctuation magnitude normalized by the average flow velocity shows non-monotonic behaviour around this value of $De_L$. So we find that the character of the spatial velocity differences is changing at a critical Deborah number.

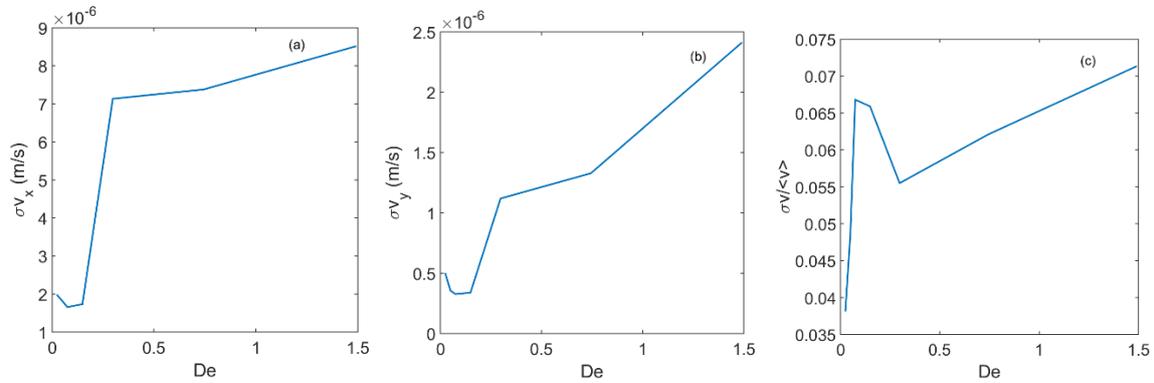

Figure 3. Spatially averaged velocity fluctuations vs. Deborah number $De_L$: (a) streamwise fluctuations, (b) lateral fluctuations, and (c) total fluctuating magnitude normalized by the average flow velocity

Figure 3 (a) and (b) show the spatially averaged velocity fluctuations along streamwise and lateral directions. Similar to figure 2 we see a sharp increase in the velocity fluctuations after a critical De number of 0.25. Figure 3 (c) shows that the magnitude of the velocity fluctuation normalized by the average flow velocity again is non-monotonic near this critical De number. So we find that even at a particular point in space, temporal fluctuations occur in the flow velocity, and that the character of these fluctuations is changing at a critical Deborah number.

The above observations show that the velocity fluctuations in the pillared microchannel are both temporal and spatial in nature, and that they change in character beyond a critical Deborah number. This is also clearly observed in time sequences of our µ-PIV images, where the onset of a flow asymmetry is clearly visible as the flow lines start to deviate from a regular laminar profile at $De_L$=0.25. When we reach $De_L$=0.90, we observe strong flow asymmetries accompanied with cross over of flow into neighboring channels (see Supplementary material). Such extreme lateral motions have never been documented before to our knowledge. Note that no such instabilities occur in Newtonian fluids at comparable flow rates.

Next we investigate the evolution of the velocity fluctuations along the channel length. To this end, we divide the whole flow domain under consideration into 100 consecutive areas, and determine the time averaged velocity fluctuations for each area.

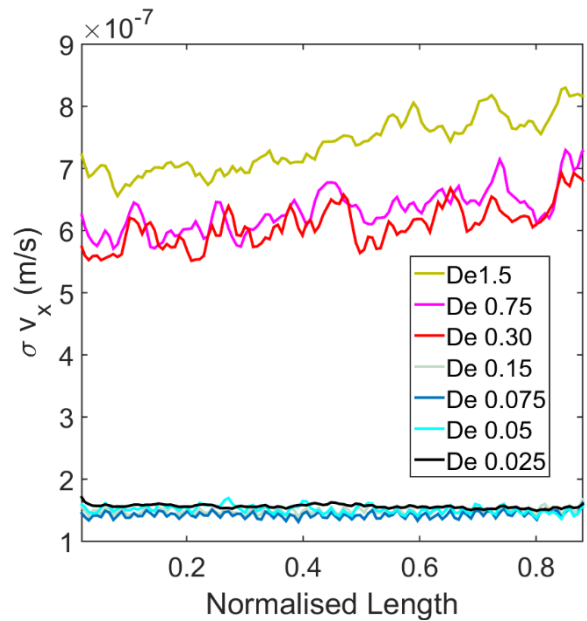

Figure 4. Velocity fluctuations as a function of position along the channel length. Different colors correspond to different Deborah numbers $(De_L)$.

Figure 4 shows the velocity fluctuations as a function of position along the channel length. Clearly, for $De_L$ larger than the critical value of 0.25 the fluctuations are large and are gradually increasing along the flow direction, while the fluctuations remain small and steady for lower Deborah numbers. The long time averaged pressure profiles obtained in our experiments also supports these velocity fluctuation observations. The critical De number, for the onset of elastic instability also agrees with the predictions of Zilz et al. [15] based on the Pakdel – Mckinley criterion [9].

The power spectrum profiles corresponding to the velocity fluctuations are shown in figure 5. We observe that both power spectra are relatively flat at lower De numbers (Newtonian regime), but shift up after a certain critical De number. This shift is most clearly visible for the lateral velocity fluctuations at low frequencies. At higher De numbers we find power law behaviour for high frequencies, with an exponent of around -3.0, which is in agreement with observations in recent literature [3, 20].

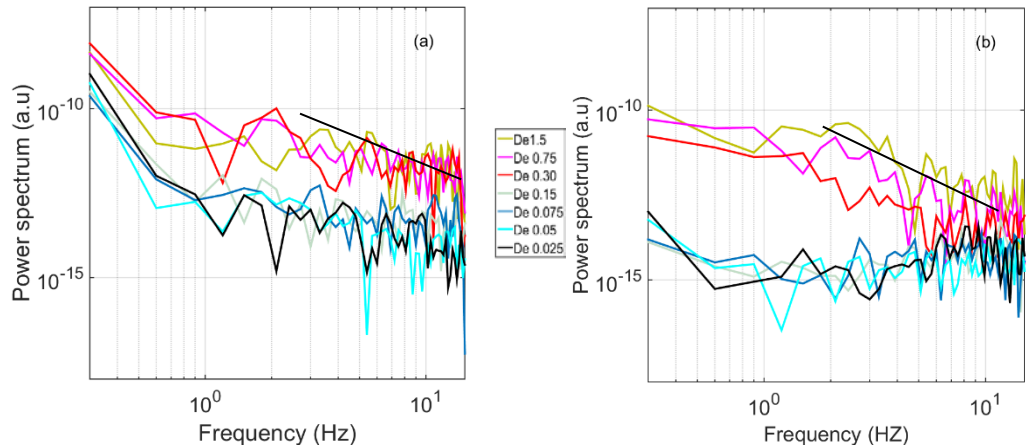

Figure 5. (a) Power spectrum of streamwise velocity fluctuations. (b) Power spectrum of lateral velocity fluctuations.

Next, we analyze the anisotropy of velocity fluctuations in the flow domain. The velocity anisotropy provides us a measure of strong flow asymmetries and presence of slow and faster moving flow channels observed at higher viscoelasticity. The ratio of the smaller to larger eigenvalue of the time-averaged velocity fluctuation tensor $\langle \delta v_x \delta v_y \rangle$ is plotted across the flow domain in Figure 6.

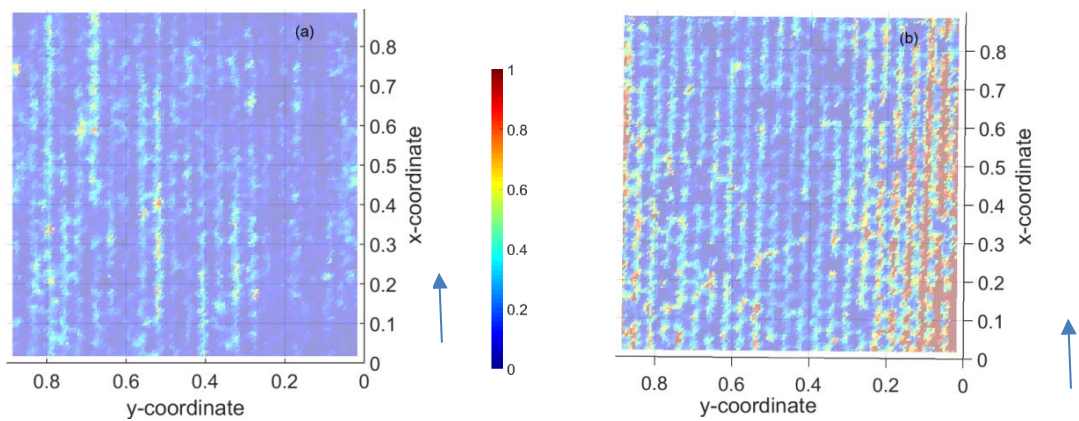

Figure 6. Flow anisotropy (normalized) in the flow domain: (a) $De_L=0.2$, (b) $De_L=1.2$ (Arrows shows flow direction in the domain)

We find that at low Deborah numbers (Fig. 6a) the velocity fluctuations are anisotropic with a typical ratio of 0.3-0.4. This is in agreement with our observations of lower lateral velocity fluctuations compared to streamwise velocity fluctuations in Fig. 2. However, at higher viscoelasticity (Fig. 6b) we see the emergence of much larger values for the eigenvalue ratio. This is in agreement with our analysis of velocity fluctuations where we observed flow asymmetry and cross-over of flow from one channel to another at higher viscoelasticity. The strong flow instability observed at higher De numbers can also be visualized by analyzing time averaged velocity vectors, shown in figure 7 and in the Supplementary videos.

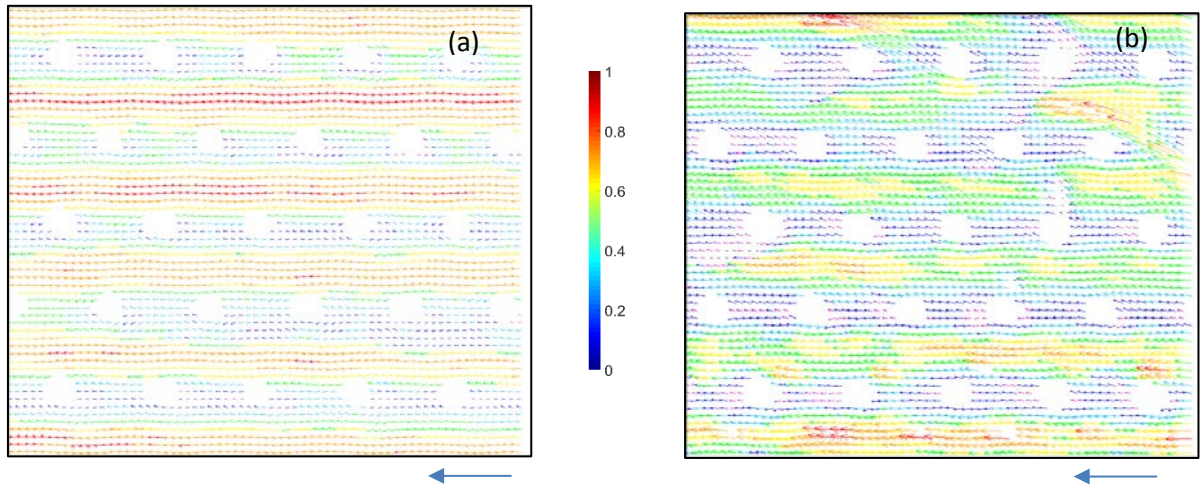

Figure 7. Time averaged velocity streamlines (normalized) at (a) $De_L$ 0.2 and (b) $De_L$ 1.0 (Arrows show flow direction in the domain)

Similar to the flow anisotropy observations we can see from figure 7 that at lower De of 0.2 the flow structure is very laminar. At De of order unity two phenomena are observed. First the presence of slow and fast moving lanes (as observed from the color contour) and sideways crossover of flow from one to another channel. Both these interesting observations can be explained as caused by elastic instabilities, if we take into account both the time scales of flow across a single cylinder ($De_P$) and across the pitch ($De_L$). According to Table 1, channel 1 has the highest confinement. At lower flow rates (<0.2 μl/min) the polymer intrinsic relaxation time is less than both these flow time scales. Hence the polymers can easily relax while flowing in between two successive pillars. At a critical flow rate of 0.2 μl/min the $De_P$ becomes of the order of 1, but $De_L$ is still less than 1.0. Thus the polymers cannot fully relax while crossing the pillars, but nevertheless they can relax in between two consecutive pillars. The local viscoelastic stresses which develop near the pillars may cause short lived instabilities, causing flow asymmetry. However, when the flow rates is more than 0.9 μl/min, both $De_P$ and $De_L$ becomes larger than 1. In that case the viscoelastic stresses become long lived, nonlinear (both spatially and temporally), and elastic turbulence sets in. This stress imbalance creates a certain flow resistance in the flow paths, forcing the polymers to change to a less resistance (sideways) path.

We note that the observed sideways crossover is non-periodic in nature and occurs far away from the walls. Also, the elastic instability is accompanied with an increase in apparent relative viscosity, defined as the ratio of pressure drop and flow rate for the viscoelastic fluid compared to that ratio for a Newtonian fluid of the same zero shear viscosity. Although from bulk rheology measurement we confirmed that our fluid is shear thinning at all measurable rates, we see an increase in apparent viscosity when Deborah is more than unity, as shown in figure 8.

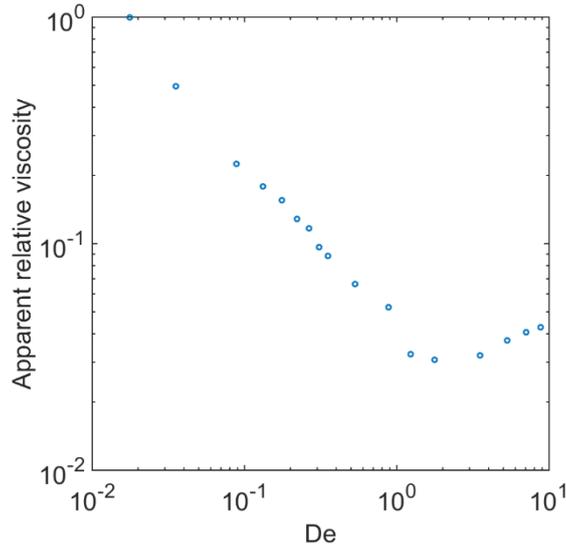

Figure 8. Plot of normalized pressure drop across the pillared channel as a function of $De_L$ number.

The continuous contractions and expansions in our geometry may cause extensional hardening of the polymer molecules. These extensional effects cause an increase in elastic stress which may drive strong spatio-temporal fluctuations, leading to instabilities. This can be proven by the fact that both the increase in pressure drop and the start of nonlinear, non-periodic instabilities occurs just after the same critical De number is reached. To verify this, we analyzed the velocity fluctuations in channel 2 and 3. We observed that the transition from laminar to first instability to elastic turbulence is directly correlated to these two Deborah numbers for channels 2 and 3, as explained for channel 1(see Supplementary material).

In summary, this experimental work shows the evidence that flow confinement has a strong effect on the development elastic instabilities in a porous medium for viscoelastic fluids. We observe very interesting flow structures with increased viscoelasticity having both temporal and spatial fluctuations, with strong crossflow motion. Moreover we observe an increase in apparent viscosity after a critical Deborah number which shows that extensional effects of polymers play a crucial role in flow with successive contraction expansions. We see these instabilities are significantly different from instabilities observed in simple shear flow, which appear at relatively larger De numbers [3] compared to our findings. Even in a model porous media with controlled conditions we observe interplay of pore configuration, rheology and elastic effects of polymer are immensely crucial for the development of flow structure. This work provides an outlook to study polymeric fluid flow through complex, random and real porous media.

This work is part of the Industrial Partnership Programme (IPP) 'Computational sciences for energy research' of the Foundation for Fundamental Research on Matter (FOM), which is part of the Netherlands Organisation for Scientific Research (NWO). This research programme is co-financed by Shell Global Solutions International B.V.